\documentclass[5p,twocolumn]{elsarticle}
\usepackage{graphicx}
\usepackage{float}
\usepackage{subfig}
\usepackage{lineno}
\usepackage{amsmath}
\usepackage{amssymb}
\usepackage{fullpage}
\journal{Nuclear Instruments and Methods A}

\usepackage{color}
\usepackage{lineno}
\usepackage{amssymb}
\usepackage{comment}
\usepackage{float}
\usepackage[normalem]{ulem}

\journal{Nuclear Instruments and Methods A}

\begin{document}
\begin{frontmatter}

\title{Nonproportionality in the scintillation light yield of bismuth germanate}

\author[a]{T. R. Gentile}
\address[a]{National Institute of Standards and Technology, Gaithersburg, MD 20899 USA}

\author[b]{M. J. Bales\corref{cor1}}
\cortext[cor1]{Corresponding author: mjbales@umich.edu (Matthew J. Bales)} 
\address[b]{University of Michigan, Ann Arbor, MI 48109 USA}

\author[c]{H. Breuer}
\address[c]{University of Maryland, College Park, MD 20742 USA}

\author[b]{T.E. Chupp}

\author[d]{K. J. Coakley}
\address[d]{National Institute of Standards and Technology, Boulder, CO 80305 USA}   

\author[e]{R. L. Cooper}
\address[e]{Indiana University, Bloomington, IN 47408 USA}

\author[a]{J. S. Nico}

\author[f]{B. O'Neill}
\address[f]{Arizona State University, Tempe, AZ 85287 USA}

\begin{abstract}

We present measurements of nonproportionality in the scintillation light yield of bismuth germanate (BGO)
for gamma-rays with energies between 6~keV and 662~keV.   The scintillation light was read out by
avalanche photodiodes (APDs) with both the BGO crystals and
APDs operated at a temperature of $\approx$~90~K.  Data were obtained
using radioisotope sources to illuminate both a single BGO crystal in a small test cryostat and a 12-element detector
in a neutron radiative beta-decay experiment.  In addition one datum was obtained in a 4.6 T magnetic field 
based on the bismuth K x-ray escape peak produced by a continuum of background gamma rays in this
apparatus.   These measurements and comparison
to prior results were motivated by an experiment to study the radiative decay  mode of the free neutron. 
The combination of data taken under different conditions yields a reasonably consistent picture for BGO nonproportionality
that should be useful for researchers employing BGO detectors at low gamma ray energies.

\end{abstract}



\end{frontmatter}

\section{Introduction}
\label{intro}

For an ideal scintillator, the number of optical photons produced by a gamma ray is linearly proportional
to the energy deposited in the scintillator by the gamma ray.  However, many scintillators
exhibit nonproportionality in their response, especially for gamma-ray energies below 100~keV~\cite{Khodyuk12}. 
Nonproportionality is typically defined by the ratio of the scintillation yield for a particular energy deposit
to the scintillation yield due to complete energy deposit for a 661.7 keV gamma ray from a $^{137}$Cs source.
Measurements of nonproportionality in the response of bismuth germanate (BGO)
in the following energy ranges have been reported:  4.5 keV to 662 keV~\cite{Averkiev90}, 6 keV to 1332 keV~\cite{Sysoeva96}, 
14 keV to 1330 keV~\cite{Moszynski04}, 60 keV to 662 keV~\cite{Verdier11}, and 11 keV to 100 keV~\cite{Khodyuk12}.
We report new measurements in the energy range from 6 keV to 662 keV.

Our measurements and comparison to prior results were motivated by an experiment to study the radiative decay
mode of the free neutron.  We reported the first observation of this process in a pilot
experiment in which gamma-rays were detected by a single BGO crystal~\cite{Nico06,Cooper10}.
To accurately test theoretical predictions for the photon energy spectrum,
a higher precision experiment has been performed with a 12-element BGO detector~\cite{Cooper12}.
Analysis of this experiment requires both Monte Carlo modeling of the
energy deposit spectrum in the BGO crystals and understanding of the light yield from BGO.
For our minimum detectable gamma ray energy deposit of $\approx$5 keV, the nonproportionality for BGO is an important source of uncertainty.
Hence we performed measurements to determine the nonproportionality observed
for our crystals, temperature, electronics, and other operating conditions.
Our results were determined by both tests on a single BGO crystal with radioactive sources
in a small test cryostat and with radioactive sources illuminating a 12-element detector
in our apparatus for neutron beta-decay.  In addition, the multi-element detector allowed for an additional measurement approach
in which for any given crystal, a continuum of background gamma rays incident on the other 11 crystals
produced escaping bismuth K x-rays that are detected by that given crystal.
This particular determination of nonproportionality at the Bi K x-ray energy (mean of 78.7 keV) was performed in a magnetic field of 4.6~T.
As in prior published studies, we determined the nonproportionality by comparing the center of the peak in the pulse
height spectrum for a given gamma-ray energy deposit $E$, $C_E$, to that expected based on linear scaling from the center 
for the 661.7 keV gamma rays from a $^{137}$Cs source, $C_\mathrm{661.7}$.  
This definition applies for sufficiently high photoelectron yields so that a symmetric peak is obtained,
which applies to all of our results except the $^{55}$Fe source results presented
at the end of the paper.  For this definition, the measured nonproportionality $N_E$ at energy $E$ expressed in percent is given by 

\begin{equation}
N_E=100~\frac{C_E}{(E/661.7)~C_\mathrm{661.7}}~\%
\label{nonprop}
\end{equation}

Hence for a ideal scintillator, $N_E$ is unity for all energies.
In Sec.~\ref{apparatus}, we describe the apparatus we employed for our measurements
of nonproportionality.  In addition to presenting our results in Sec.~\ref{Results}, we compare to previous reports.
The combination of data taken under different conditions yields a consistent picture for BGO nonproportionality
that should be useful for researchers employing BGO at low gamma ray energies.
In Sec.~\ref{Fedata} we present results for data obtained in the test cryostat 
with 6 keV x-rays from an $^{55}$Fe source and in Sec.~\ref{concl} we summarize the paper.

\section{Apparatus and measurements}
\label{apparatus}

\subsection{Test cryostat measurements}

\begin{figure*}
\begin{center}
\includegraphics[width=\textwidth]{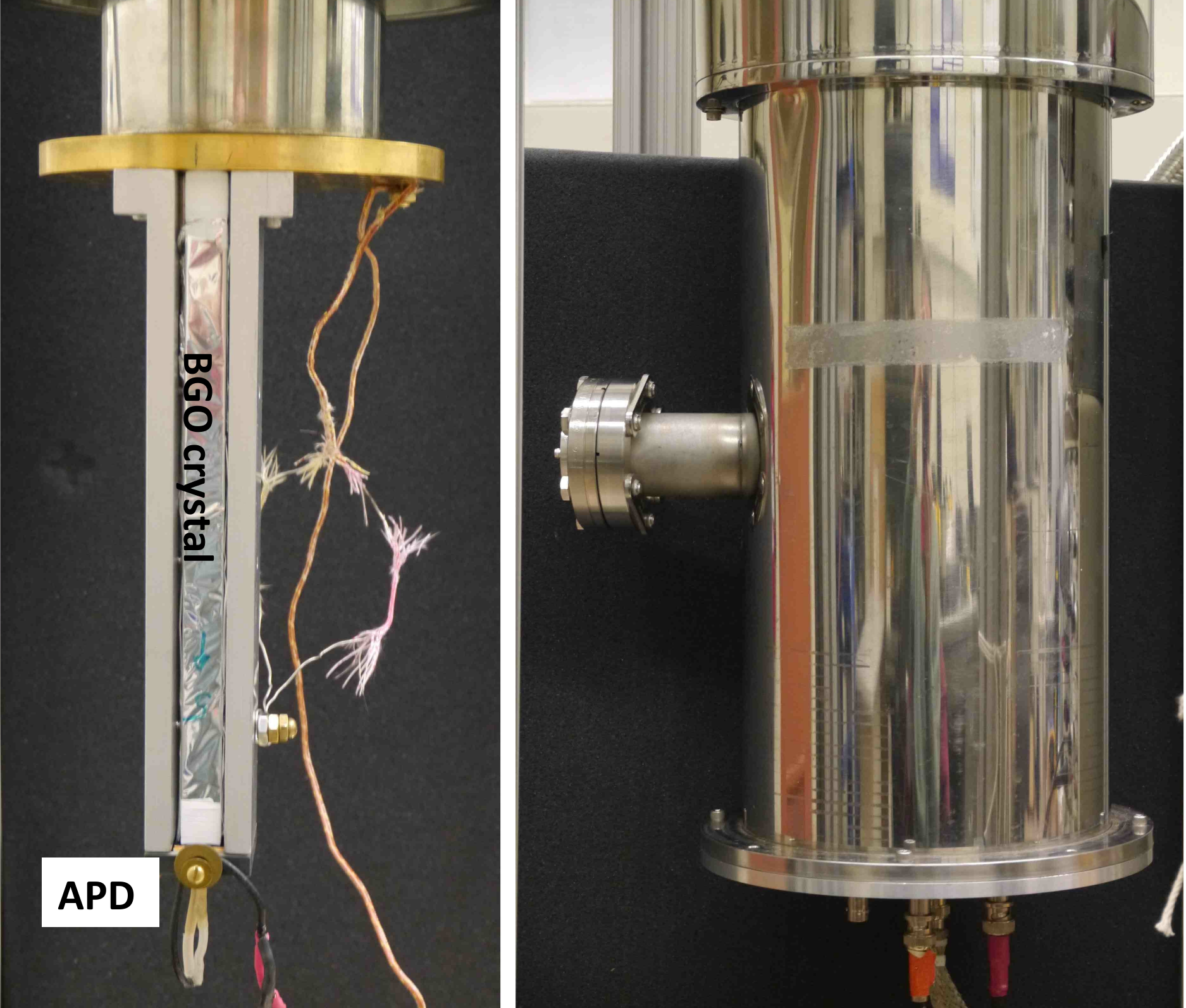}
\caption{Photographs of test cryostat apparatus.  The left photograph shows the crystal and APD
with the cryostat vacuum jacket and radiation shield removed.  At the top of the photograph
the inner reservoir of the dewar is visible.  In the right photo
(taken with cryostat rotated by 90 degrees)
the cryostat is closed up and the vacuum port for illumination of the center of the crystal
with sources located outside the cryostat vacuum is shown. (Color online only.)}
\label{fig:photos}
\end{center}
\end{figure*}

The test cryostat apparatus has been previously described in its use for studies of APD response~\cite{Gentile12}.
For these nonproportionality studies, a 12 mm by 12 mm by 200 mm BGO crystal \cite{Rexon}
coupled to an APD with an active area of 13.5 mm by 13.5 mm~\cite{RMD} was thermally sunk  to the bottom of the 
inner reservoir of a cryostat (see Fig.~\ref{fig:photos}).  As discussed in Ref.~\cite{Cooper12}, the opposite end of the BGO crystal was painted
with reflective paint to improve the light collection efficiency.   The long sides of the BGO crystal were covered by two layers of aluminized Mylar
(total of 12.4 $\mu$m Mylar and 0.065 $\mu$m aluminum). 
The long-side regions near the end faces ($\approx$12 mm) had additional 
layers of polytetrafluoroethlyene (PTFE) tape and aluminized Mylar for mechanical protection. 
($\approx$300 $\mu$m PTFE and 6 $\mu$m Mylar with 0.033 $\mu$m aluminum).
The cryostat's vacuum jacket included a 3.5 cm diameter tube with a 7 cm diameter metal-sealed flange through which
we illuminated  the crystal about midway along its 200 mm length (see right side of Fig.~\ref{fig:photos}).
The light collection efficiency for gamma rays incident at different locations along this crystal can vary by $\approx$~8~\% from the center, with greater light collection efficiency obtained at either end of the crystal.  This geometric effect is independent of the incident energy~\cite{Cooper12}.  For these nonproportionality tests, the tubing for the vacuum port and the cryostat itself provided sufficient collimation to ensure that only the central region of the crystal was illuminated. For the highest energy (thus most penetrating) source, this was confirmed by a check with additional  collimation
that consisted of 1.2 cm thick, 2.8 cm tall copper pieces clamped in place above and below the horizontal 3.5 cm tube.

An oil-free turbomolecular pumping station was used to reduce the pressure in the chamber to a typical
base pressure of on the order of 10$^{-6}$ mbar.   Upon filling the cryostat with
liquid nitrogen, the typical pressure was on the order of 10$^{-7}$ mbar.
The APD signal was sent to a charge-sensitive preamplifier\cite{Canberra},
followed by an amplifier set for a gain of 100 and a 6~$\mu$s shaping time constant. 
The output of the shaping amplifier was registered by a multichannel analyzer (MCA)\cite{PGT}.
The lower level discriminator of the MCA was typically set 
slightly below the detector noise level  at $\approx$ 10 channels, for a 1000-channel full range of $\approx$1000~keV.
The APD was operated at a typical bias of 1380~V ($\approx$40~V below breakdown).  
Using studies with a precision pulser, we found that our electronics
for the test cryostat apparatus exhibited a slight nonlinearity that was consistent with an offset of one channel out of 1000 in the MCA.
To determine the BGO nonproportionality, we corrected for this nonlinearity, which yielded a 
shift of 4.3~\% for our lowest energy datum at 23.3 keV, decreasing to 1~\% at 60 keV.

A thin polyimide window in the side port allowed the sources to be outside the
cryostat vacuum space so they could be easily manipulated.  The gamma-ray energies and sources
employed in the test cryostat apparatus are listed in Table~\ref{table:sources}.
A $^{137}$Cs source was used to provide the 662 keV reference.
The mean energy deposit listed was determined for multiple-line emissions from the following components in keV, using the 
weighting percentages~\cite{nuclidetables} in parentheses:  $^{109}$Cd - 22.1, 25.1 (83.7*0.026, 17.7*0.078);
$^{133}$Ba - 30.6, 31.0, 35.1, 35.9 (54, 100, 29, 7); $^{153}$Gd - 97.4, 103.2 (29, 21); and $^{57}$Co -
122.1, 136.5 (85.5, 10.7).  For the low energy lines of the $^{109}$Cd source, the weighting included
the factors shown to account for energy-dependent transmission through a 0.25 mm thick stainless steel cover on
this particular source.  To evaluate the nonproportionality, the pulse height spectrum for a given emission
was fit with a single Gaussian, except for the multiple-line emissions noted above for which multiple-line Gaussians were employed.
In the case of multiple-line emission, a single nonproportionality value was used for all the Gaussians
and is listed for the mean energy deposit.

\begin{table}
\begin{center}
  \begin{tabular} {cccc}
    \hline
    $E$ [keV] & source & $N_E$ [\%]  \\ 
    \hline
        23.3 & $^{109}$Cd &  73.4 (3.0)  \\ 
        31.7 & $^{133}$Ba &  81.2 (2.5)  \\
        59.5 & $^{241}$Am & 88.8 (1.0)\\
        81.0 & $^{133}$Ba & 91.6 (1.5) \\
        88.0	& $^{109}$Cd & 90.7 (1.0) \\
        99.8 & $^{153}$Gd & 86.5 (1.5)  \\
        123.7 & $^{57}$Co & 89.2 (1.5)  \\
        185.7 & $^{235}$U & 93.3 (2.0) \\
        661.7 & $^{137}$Cs & 100.0 \\     
        \hline
  \end{tabular}
   \end{center}
  \caption {Mean gamma-ray energy deposits and nonproportionality $N_E$ for the sources employed with
the test cryostat apparatus.  For sources with multiple, unresolved emission lines, the first column shows the mean energy $E$ deposit
(see text).  The numbers in parentheses show the one-sigma uncertainty in the last digits, which were estimated
based primarily on reproduceability and uncertainty in the determination of peak locations.
The nonproportionality is defined to be 100~\% for the photopeak produced by the 661.7 keV gamma ray from $^{137}$Cs. }
\label{table:sources}
\end{table}

\begin{figure}
\begin{center}
\includegraphics[scale=0.5]{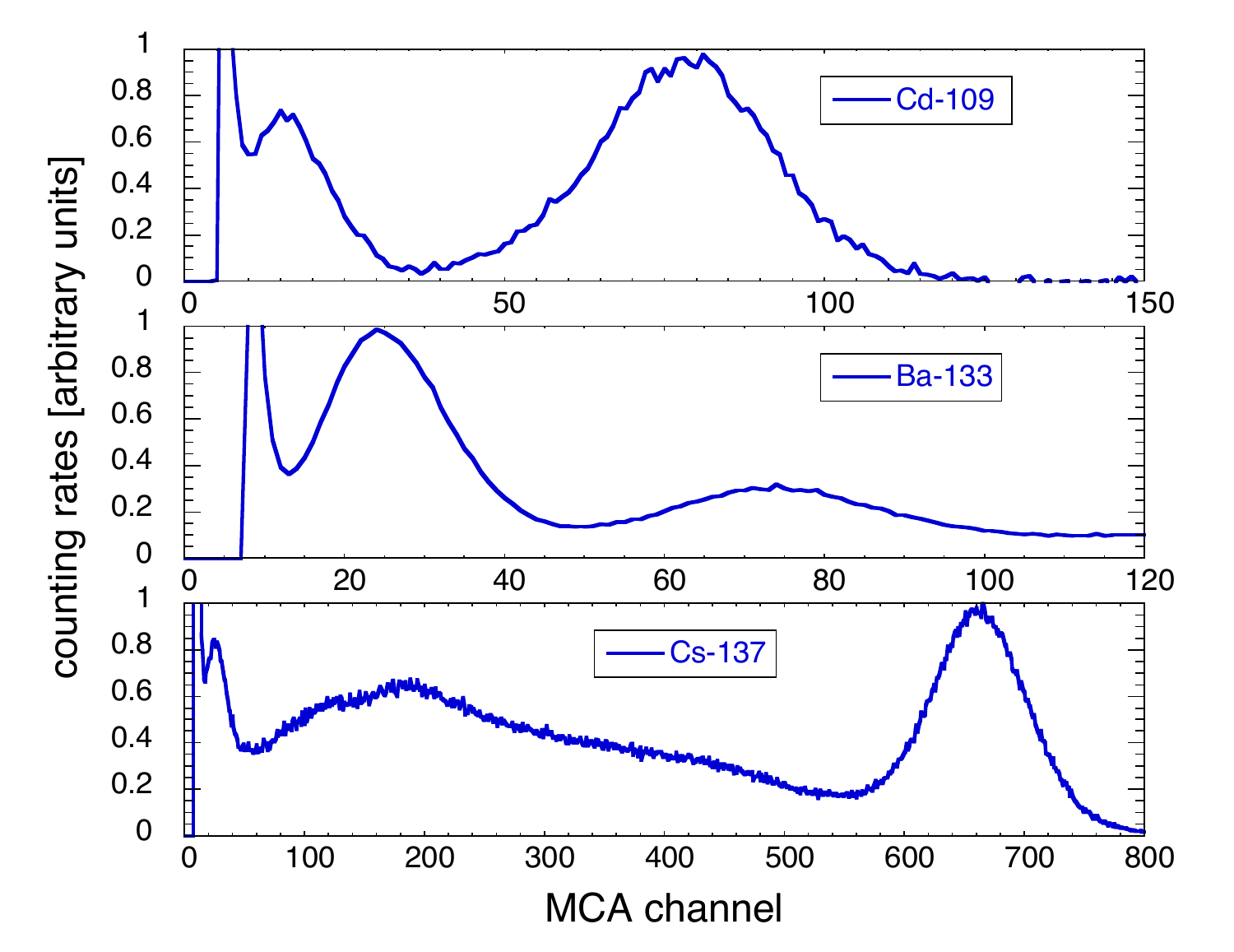}
\caption{Pulse height distributions obtained in the test cryostat apparatus for three of the seven sources employed
($^{109}$Cd, $^{133}$Ba, and $^{137}$Cs).  The resolution is primarily determined by photoelectron statistics.}
\label{fig:spectra}
\end{center}
\end{figure}

Fig.~\ref{fig:spectra} shows the background-subtracted pulse height distributions for three of the seven sources.  
The background was determined by counting without the source.  However, we found that the apparent threshold
was higher with a source than without a source and it varied between sources.  The response observed near the 10 keV threshold is 
due primarily to this source-related response, with a smaller contribution from imperfect background subtraction.

\subsection{Measurements in neutron beta-decay apparatus}

The apparatus for neutron beta-decay has been described elsewhere~\cite{Nico06,Cooper10,Cooper12}.
In brief, a neutron beam passes through the cylindrical bore of a superconducting magnet
and gamma rays emitted during neutron beta decay are detected by an annular array
of twelve 12 mm by 12 mm by 200 mm  BGO crystals. 
As in the test dewar, each crystal is coupled to an APD with an active area of 13.5 mm by 13.5 mm.
However, the electronic processing and spectrum analysis were different than
that performed for the test dewar studies.
The preamplifier output signals were directly fed into two 8-channel GaGe
cards~\cite{Gage}, which were set to digitize the signals with 40 ns per channel resolution
in 80 $\mu$s long traces.  The decay time constant of the preamplifiers is 50 $\mu$s.
The amplitude of the digitized signals (see Ref.~\cite{Cooper12} for an example)
were determined by fitting the signals to a template waveform.
The template was a model waveform constructed from the average of  
preamplifier signals for gamma rays with a range of energies.  

To test the detector response with radioactive sources, we employed a reentrant tube
that allowed the sources to be outside the magnet's vacuum chamber while illuminating
the liquid nitrogen-cooled detector in a similar way as the photons from neutron decay.
These tests were performed with the apparatus off the neutron beam line and with the magnetic field off.
Pulse height spectra  with $^{137}$Cs, $^{133}$Ba, $^{57}$Co, and $^{241}$Am  sources
are shown in Ref.~\cite{Cooper12}, along with the results of modelling
the detector response based, in part, on MCNP5~\cite{Brown03} predictions for the energy deposit spectrum
in each  BGO crystal.   Nonproportionality was not included in the modeling.
For this paper, we determined nonproportionality by comparing the location of emission lines
from sources to those determined from Geant4 \cite{geant4ref} and MCNP5 modelling
with the assumption that the nonproportionality was 100~\% for all energy deposits.   A $^{137}$Cs source was used to provide the
662 keV reference line for lines from a $^{133}$Ba (nominally 31 keV, 81 keV, and 356 keV)
and a $^{57}$Co source (122 keV and an X-ray escape peak at 43 keV).

An additional datum was obtained using x-ray escape from continuum background gamma-rays detected
during operation on the neutron beam line.  Although nonproportionality may depend
weakly on the energy deposition process\cite{Khodyuk10}, in Ref.~\cite{Moszynski04} the values obtained 
with escape peaks were the same within uncertainties as those obtained from direct sources.
For a given BGO crystal, escape of Bi K x-rays
from all the other crystals yields a peak at a mean energy of 78.7~keV.
If the pulse height spectra from all the crystals are summed, this peak disappears
because detection of the x-ray only occurs when the x-ray escaped from a different crystal.
To improve the extraction of the peak location, we combined all the data from the 12 crystals
but also subtracted off the summed spectrum.
In this case, the magnetic field was 4.6~T.  The gamma-ray energies and sources
employed in the neutron radiative decay (RDK) apparatus are listed in Table~\ref{table:RDKsources}.

\begin{table}
\begin{center}
  \begin{tabular} {cccc}
    \hline
    $E$ [keV] & source & $N_E$ [\%]  \\ 
    \hline
        32.3 & $^{133}$Ba &  81.0 ( 2.0) \\ 
        42.7 & $^{57}$Co (escape) & 87.0 (2.0) \\
        78.7 & Bi escape & 94.0 (2.0) \\
        79.7 & $^{133}$Ba & 93.0 (1.0) \\
        121.9 & $^{57}$Co & 92.0  (1.0) \\
        356.0 & $^{133}$Ba & 98.6 (0.2) \\
        661.7 & $^{137}$Cs & 100.0 \\     
        \hline
  \end{tabular}
  \end{center}
  \caption {Mean gamma-ray energy deposits and nonproportionality $N_E$ for the sources employed 
with the neutron beta-decay apparatus.  The first column shows the mean energy $E$ deposit
as determined from our Monte Carlo modeling of the apparatus, without including nonproportionality.
(Note that these values are not necessarily exactly the same as those in Table~\ref{table:sources}
because of a different experimental configuration and different methods for their determinations.)
The nonproportionality is defined to be 100~\% for the 661.7 keV gamma ray from $^{137}$Cs.
The numbers in parentheses show the one-sigma uncertainty in the last digits, which were determined from uncertainty in the
determination of peak locations and the variance in the expected peak location from the Monte Carlo for the 11 BGO crystals.}
\label{table:RDKsources}
\end{table}

\section{Results} 
\label{Results}

Information on our results and those from past literature are summarized in Table~\ref{table:identification}.
Whereas most results were obtained with a variety of radioisotopes, the work in Ref.~\cite{Khodyuk12}
employed monochromatic gamma-ray beams from a synchrotron with resolutions varying from 2 eV at 9 keV 
to 20 eV at 100 keV.  
In Ref.~\cite{Moszynski04} a photomultiplier tube (PMT) was used for measurements at room temperature and
an APD~\cite{Adv} was used for measurements near liquid nitrogen temperature, with similar results being obtained
for the nonproportionality.  In addition to direct gamma ray lines, nonproportionality was determined
using escape peaks, again with similar results.  
 
\begin{table}
\begin{center}
  \begin{tabular} {ccccc}
    \hline
    ID & source & $T$~[K] & opdet & crystal [mm]  \\ 
    \hline
       & \cite{Averkiev90}, R & 295 & PMT &  not stated   \\
       & \cite{Sysoeva96}, R & 295 & PMT & 10-20, 1-5   \\
    1 & \cite{Moszynski04}, R & 295 & PMT & 9, 4 or 9   \\
    2 & \cite{Moszynski04}, R & 100 & APD & 9, 4    \\
    3 & \cite{Moszynski04}, ER & 100 & APD & 9, 4    \\
      &  \cite{Verdier11}, R & 3, 10, 30 & PMT &20x10x5 \\
    4 & \cite{Khodyuk12}, S & 295 & PMT & 5-10, 1   \\
    5 &  test, R & 90 & APD & 12x12x200    \\
    6 & RDK, R & 90 & APD & 12x12x200    \\   
    7 & RDK, E & 90 & APD & 12x12x200   \\  
        \hline
  \end{tabular}
  \end{center}
  \caption {Identification (ID) for numerical references in Fig.~\ref{fig:nonlindata} and information
on experimental conditions from the literature.  The information for our data obtained using the test cryostat (test) and the
neutron beta-decay apparatus (RDK) are reported.
We list the source of the gamma rays, the approximate temperature $T$ of the BGO crystal, 
the method for readout of the optical photons (heading \textquotedblleft opdet\textquotedblright ), and the crystal
dimensions in millimeters (heading \textquotedblleft crystal\textquotedblright) .
The source are synchrotron radiation (S), radioisotopes (R), escape peaks from radioisotopes (ER), and 
Bi escape X-rays produced from a continuum of background gamma rays in our neutron beta-decay apparatus (E).
For cylindrical crystals the dimensions listed are in the form:  diameter, length.}
\label{table:identification}
\end{table}

\begin{figure*}
\begin{center}
\includegraphics[scale=1]{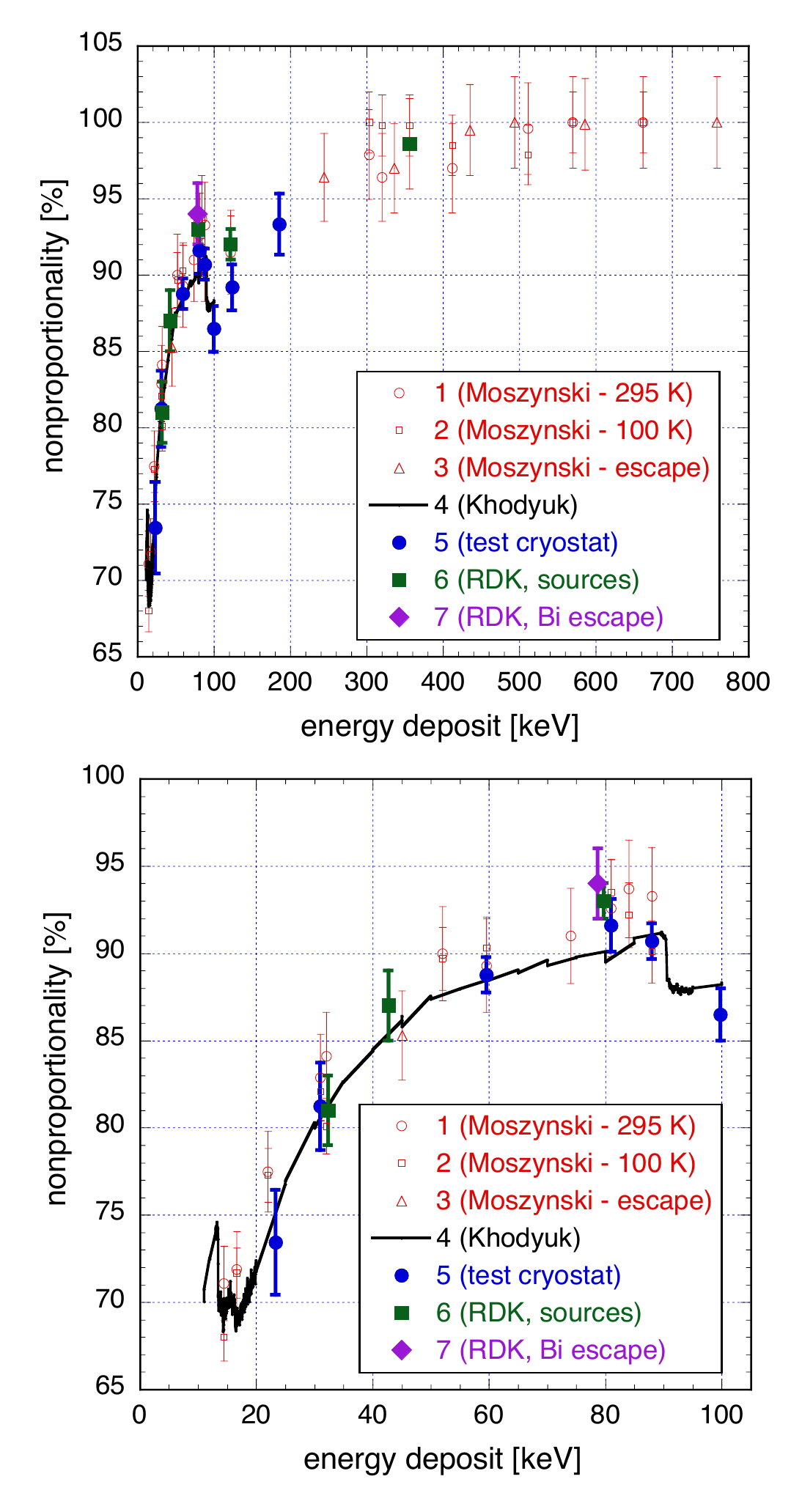}
\caption{Data for nonproportionality in the response of BGO in the gamma-ray energy range 
from (a) 0 keV to 800~keV and (b) 0 keV to 110 keV.   See Table~\ref{table:identification} for identification and information on each data set.
For clarity, the nearly continuous and precise (0.5~\% uncertainties) synchrotron data from Ref.~\cite{Khodyuk12}
are shown as a line with small points.   The error bars for our test dewar and RDK apparatus data
correspond to the uncertainties listed in Table~\ref{table:sources} and Table~\ref{table:RDKsources}, respectively.}
\label{fig:nonlindata}
\end{center}
\end{figure*}

Fig.~\ref{fig:nonlindata} shows data for nonproportionality in the response of bismuth germanate (BGO).
Our results from the test cryostat and neutron beta-decay apparatus are listed
in Tables~\ref{table:sources} and \ref{table:RDKsources}, respectively.
The combination of our data and prior published results include a variety of crystal sizes and sources, 
optical readout methods, temperatures, radiation sources, and analysis methods.
Nevertheless, all of the results generally agree within a few percent.
The results from our test cryostat data are closest to the results of Ref.~\cite{Khodyuk12},
whereas the results from Ref.~\cite{Moszynski04} and our RDK apparatus are 
typically a few percent higher, in particular just below the Bi K-edge at $\approx$90~keV.  The single datum obtained
at the bismuth K x-ray energy at a 4.6 T magnetic field is consistent within uncertainties
with the nonproportionality values obtained at near zero magnetic field.

\section{Nonproportionality at 6 keV}
\label{Fedata}

\begin{figure}
\begin{center}
\includegraphics[scale=0.45]{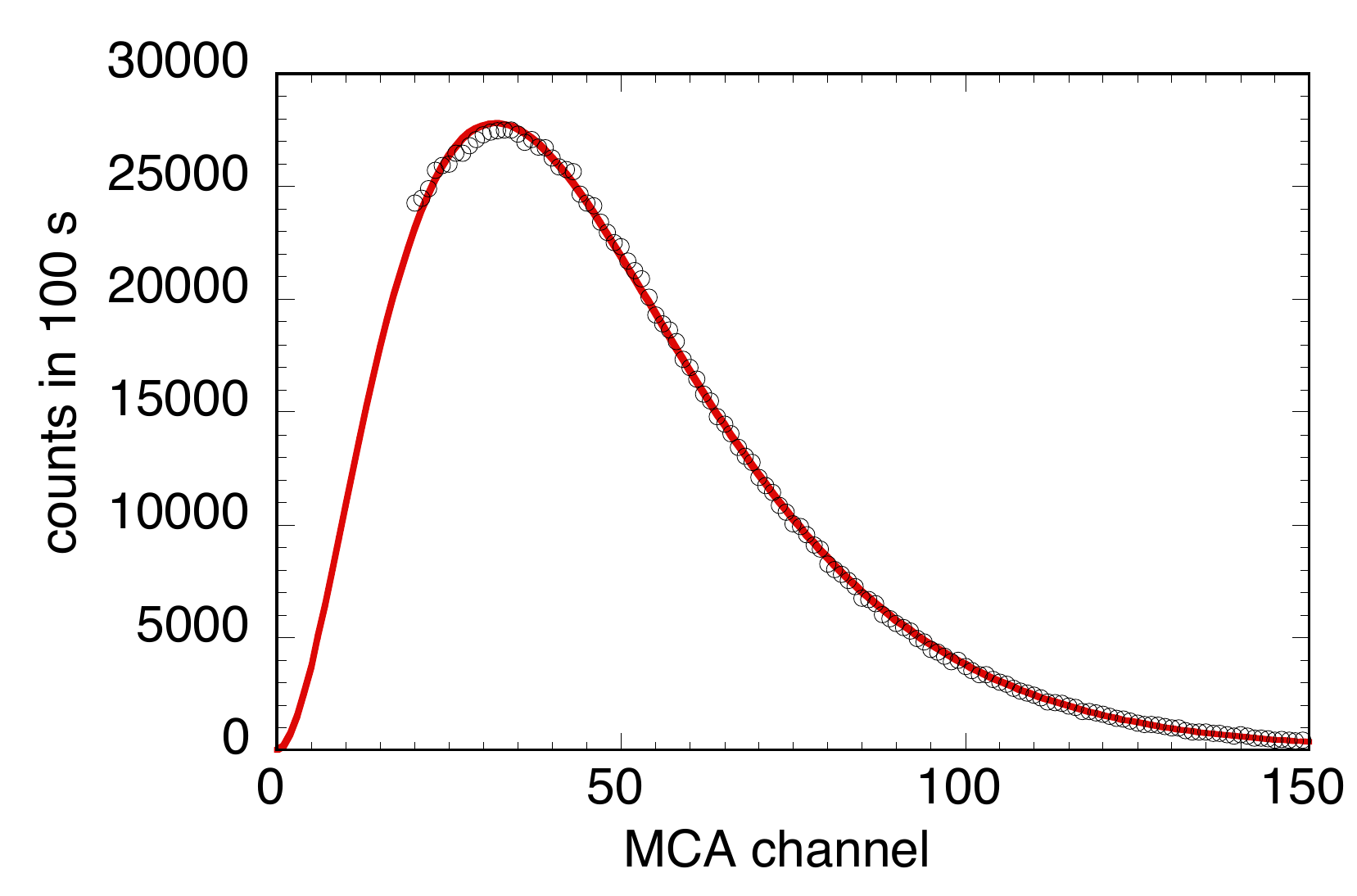}
\caption{Pulse height distribution obtained in the test cryostat apparatus for 1 cm$^3$ BGO crystal
illuminated by an $^{55}$Fe source.  The solid line is a fit to a gamma distribution.}
\label{fig:Fedata}
\end{center}
\end{figure}

Ref.~\cite{Averkiev90} reported a monotonic increase in $N_E$
with decreasing energy deposits in the range below the Bi L edge at $\approx$13~keV.
Whereas the expected increase in the vicinity of the L-edge was reported
in Refs.~\cite{Khodyuk12} and \cite{Sysoeva96}, $N_E$ was found to continue decreasing further below the edge region.
This discrepancy motivated us to extend our nonproportionality measurements as low as possible in photon energy,
hence we illuminated a 1 cm$^3$ BGO crystal with 6 keV x-rays from an $^{55}$Fe source with an activity of $\approx$9 mCi.
We obtained the best signal to noise ratio for observing these low amplitude pulses
by operating the APD at between 5 V and 15 V below breakdown.
For these data, the amplifier gain was 116, the shaping time constant was 6~$\mu$s, and the lower level discriminator
of the MCA was 13 channels, for a 1000-channel full range of $\approx$100~keV.
Under these conditions, we obtained the background-subtracted pulse-height spectrum shown in Fig.~\ref{fig:Fedata}.
To avoid extremely low channels on the MCA, the nonproportionality was determined with respect to the 59.5~keV peak
from the $^{241}$Am source and a correction was applied based on our nonproportionality value for this line.
Because of the peak asymmetry, we fit the data to a gamma distribution\cite{Gale66} (see Fig.~\ref{fig:Fedata}).
Evaluation of the mean value yielded $N_{6.0}= (69\pm3)~\%$,
which corresponds to 1.4 times greater than that would be obtained based on the observed peak location.

An asymmetric spectrum would be expected from Poisson statistics for a yield of only a few photoelectrons or less.
In this case APD noise might mask the appearance of resolved peaks for specific photoelectron numbers. 
We modelled the pulse height spectrum using Poisson statistics and obtained the best results for a photoelectron yield of 2.6.
However, an estimate of the photoelectron yield is not consistent with such a low value. 
For our large crystals we had estimated the photoelectron yield to be $\approx$1000 for 662 keV gamma rays\cite{Cooper12}.
For 6 keV and $N_{6.0}=100$~\% this estimate corresponds to 9 photoelectrons, but the observed nonproportionality
reduces the estimated yield to 6 photoelectrons. In addition, the small crystal employed for the $^{55}$Fe studies
provided a higher yield.  Hence the substantial asymmetry observed might indicate a pulse height distribution with a variance
greater than its mean, due to APD noise.  The impact of the value of the yield
is on the probability of zero photoelectrons, which yields a small correction to the observed nonproportionality. 
Based on our expected yield the correction is negligible. However, we cannot discount the possibility
that the yield is $\approx$2.6 photoelectrons, which would result in a probability of zero photoelectrons of $\exp(-2.6)$=0.074.
From these considerations we assign an asymmetric one-sigma uncertainty interval, hence our final value
for the nonproportionality at 6.0 keV is $N_{6.0}=(69^{+3}_{-8})~\%$.
Our result falls midway between the values reported in Refs.~\cite{Khodyuk12} and \cite{Sysoeva96}. 
In Ref.~\cite{Sysoeva96} it was suggested that the difference between their results and those of Ref.~\cite{Averkiev90} could be related
to the surface state of the investigated crystals.


\section{Conclusions}
\label{concl}

Motivated by an experiment to study neutron radiative beta-decay, we performed
new measurements of the nonproportionality in the light yield of BGO.
For the energy range from 23 keV to 662 keV, these measurements were obtained in two different apparatus
by illuminating long crystals at a temperature of $\approx$~90~K with gamma rays from radioisotope sources
and reading out the scintillation light with APDs.  In addition one datum was obtained in a 4.6 T magnetic field 
based on the bismuth K x-ray escape peak produced by a continuum of background gamma rays.    
The combination of data taken under different conditions yields a reasonably consistent picture for BGO nonproportionality.
An additional measurement using 6~keV X-rays from an $^{55}$Fe source was obtained with a small crystal.
An asymmetric peak was observed,
hence a value for the nonproportionality was obtained by extracting the mean value for
a fit to a gamma distribution.  

\vspace {0.2 in}

{\bf {\noindent Acknowledgements}}

We thank I.V. Khodyuk and M. Moszynski for providing their numerical data for 
incorporation in the figures, and for their comments.  We also thank Larry Lucas
and Janna Shupe for loan of the $^{109}$Cd, $^{153}$Gd, and $^{235}$U sources.
We acknowledge support from the Department of Energy, Office of Nuclear 
Physics, and the National Science Foundation.


\begin{thebibliography}{200}

\bibitem{Khodyuk12}  I.V. Khodyuk and P. Dorenbos, IEEE Trans. Nucl. Science {\bf 59}, 3320 (2012).

\bibitem{Averkiev90}  V.V. Averkiev, V.K. Lyapidevskii and G. Kh. Salakhutdinov, 
Pribory i Tekhnika Eksperienta {\bf 4}, 80 (1990).

\bibitem{Sysoeva96} E.P. Sysoeva, O.V. Zelenskaya and E.V. Sysoeva, IEEE Trans. Nucl. Science {\bf 43}, 1282 (1996).

\bibitem{Moszynski04}  M. Moszynski {\it et al.}, IEEE Trans. Nucl. Science {\bf 51}, 1074 (2004).

\bibitem{Verdier11} M.-A. Verdier, {\it et~al.}, Phys. Rev. B {\bf 84}, 214306 (2011) 

\bibitem{Nico06} J.S. Nico {\it et al.}, Nature {\bf 444}, 1059-1062 (2006). 

\bibitem{Cooper10}  R.L. Cooper {\it et al.}, Phys. Rev C {\bf 81}, 035503 (2010). 

\bibitem{Cooper12}  R.L. Cooper {\it et al.}, Nucl, Instrum. Meth. A {\bf 691}, 64 (2012).  

\bibitem{Gentile12}  T.R. Gentile, M. Bales, U. Arp, B. Dong, and R. Farrell, Rev. Sci. Instrum. {\bf 83}, 053105 (2012). 
 
 \bibitem{Rexon}  Rexon Components, Inc., 24500 Highpoint Rd., Beachwood, OH 44122.
 Certain trade names and company products are mentioned in the
text or identified in an illustration in order to
adequately specify the experimental procedure and equipment used.  In
no case does such identification imply recommendation or endorsement
by the National Institute of Standards and Technology, nor does it
imply that the products are necessarily the best available for the purpose.
 
\bibitem{RMD}  Radiation Monitoring Devices, Inc., Watertown, MA 02472.


\bibitem{Canberra} Model 2006, Canberra Industries, 800 Research Parkway, Meriden, Connecticut, 06450.

\bibitem{PGT}  Quantum MCA8000, Princeton Gamma-Tech, Inc., C/N 863, 
Princeton, NJ 08542-0863.

\bibitem{nuclidetables}  http://www.nucleide.org/DDEP\_WG/Nuclides/Cd-109\_tables.pdf,
and similarly for other sources.  

\bibitem{Gage}
Gage CompuScope 8289 (Octopus 12-bit CompuScope A/D card: 125 MS/s per channel,
  8 channels, 128 MS memory), DynamicSignals LLC, 900 N.\,State St., Lockport,
  IL 60441.

\bibitem{geant4ref}  S. Agostinelli {\it et al.}, Nucl., Instrum. Meth. A {\bf 506} 250 (2003).

\bibitem{Khodyuk10}  I.V. Khodyuk, J.T.M. de Hass, and P. Dorenbos, IEEE Trans. Nucl. Science {\bf 57}, 1175 (2010).

\bibitem{Brown03}  F.~B. Brown, {\it et~al.}, A General Monte Carlo N-Particle Transport Code, Version 5
  LA-UR-03-1987 (2003).

\bibitem{Adv} 16 mm diameter APD, Advanced Photonix, 1240 Avenida Acaso, Camarillo, CA 93012.

\bibitem{Gale66}  H.J. Gale and J.A.B. Gibson, J. Sci. Instrum. {\bf 43}, 224 (1966).


%
%
%
%
%
%

%


\end{thebibliography}
\end{document}